\documentclass[sigconf]{acmart}

\usepackage{bm}
\usepackage{booktabs} % For formal tables
\usepackage{graphics}
\usepackage{multirow}
\usepackage{flushend}
\usepackage[english]{babel}

\begin{document}

%Conference
\copyrightyear{2018}
\acmYear{2018}
\setcopyright{acmlicensed}
\acmConference[CIKM '18]{The 27th ACM International Conference on
Information and Knowledge Management}{October 22--26,
2018}{Torino, Italy}
\acmBooktitle{The 27th ACM International Conference on Information
and Knowledge Management (CIKM '18), October 22--26, 2018, Torino,
Italy}
\acmPrice{15.00}
\acmDOI{10.1145/3269206.3271728}
\acmISBN{978-1-4503-6014-2/18/10}

\title{Personalizing Search Results Using Hierarchical RNN \\with Query-aware Attention}

\author{Songwei Ge$^{1,3}$, Zhicheng Dou$^{1,2,3}$, Zhengbao Jiang$^{1,2}$, Jian-Yun Nie$^5$, and Ji-Rong Wen$^{1,2,4}$}
\affiliation{%
  $^1$School of Information, Renmin University of China, $^5$DIRO, Universit\'{e} de Montr\'{e}al
}
\affiliation{$^2$Beijing Key Laboratory of Big Data Management and Analysis Methods}

\affiliation{$^3$National Engineering Laboratory of Big Data System Software ( Beijing Institute of Technology )}

\affiliation{$^4$Key Laboratory of Data Engineering and Knowledge Engineering, MOE}

\email{gesongwei666@gmail.com, dou@ruc.edu.cn, rucjzb@163.com, nie@iro.umontreal.ca,jirong.wen@gmail.com}

\begin{abstract}
Search results personalization has become an effective way to improve the quality of search engines. Previous studies extracted information such as past clicks, user topical interests, query click entropy and so on to tailor the original ranking. However, few studies have taken into account the sequential information underlying previous queries and sessions. Intuitively, the order of issued queries is important in inferring the real user interests. And more recent sessions should provide more reliable personal signals than older sessions. In addition, the previous search history and user behaviors should influence the personalization of the current query depending on their relatedness. To implement these intuitions, in this paper we employ a hierarchical recurrent neural network to exploit such sequential information and automatically generate user profile from historical data. We propose a query-aware attention model to generate a dynamic user profile based on the input query. Significant improvement is observed in the experiment with data from a commercial search engine when compared with several traditional personalization models. Our analysis reveals that the attention model is able to attribute higher weights to more related past sessions after fine training.
\end{abstract}


\settopmatter{printacmref=false, printfolios=false}

\maketitle

% \vspace{-0.08cm}
{\fontsize{7.9pt}{7.9pt} \selectfont
\textbf{ACM Reference Format:}\\
Songwei Ge, Zhicheng Dou, Zhengbao Jiang, Jian-Yun Nie, and Ji-Rong Wen. 2018. Personalizing Search Results Using Hierarchical RNN with Query-aware Attention. In The 27th ACM International Conference on Information and Knowledge Management (CIKM'18), October 22--26, 2018, Torino, Italy. ACM, New York, NY, USA, 10 pages. https://doi.org/10.1145/3269206.3271728
}

% \vspace{-0.08cm}
\section{Introduction}
\label{sec:intro}
Users come to search engine with specific intents, but the queries they issue often fail to express accurate meanings. It is often the case that users issue the same query to express different search intents. Therefore, returning the same results to all users is not the best strategy for a search engine. Personalization potentially solves this problem: by appropriately modeling user interest profiles, personalized search is able to alleviate the ambiguity problem by providing more precise results to individual users. 

Personalizing search results to particular users based on their profiles has been a hot field in both academia \cite{bennett2012modeling, cai2014personalized, song2014adapting, teevan2011understanding} and industry \cite{hannak2013measuring} for a long time. Among these studies, many personalization methods utilized click-through data to generate user profiles, since large-scale query logs contain strong indications on users' preferences and are also relatively cheap to acquire when compared with manual labels \cite{bennett2012modeling, joachims2002optimizing, song2014adapting, speretta2005personalized,vu2015temporal}. These studies demonstrated that through mining personalized signals from query logs, a search engine can re-rank search results and obtain more relevant rankings for individual users. In this paper, we focus on exploiting information from the sequences of past queries and sessions and build a dynamic user profile based on input query. 

Previous studies have revealed that diverse features, such as click counts, user topical interests, click entropy and so on, could be extracted from user query logs to help personalize search results. \cite{dou2007large, bennett2012modeling, harvey2013building, sontag2012probabilistic, Vu2014Improving}. However, few studies have leveraged sequential information hidden in past queries and sessions. Intuitively, a more recent query or session should contribute more than an older one to the current search. For example, a user may have searched for "cherry blossom Japan", "cherry jam" and so on, but is now interested in mechanical keyboard and intends to obtain some introductions. His query "cherry reviews" should be more related to the keyboard brand rather than the fruit or the flower. Previous studies attempted to simply apply an exponential decay to distinguish the recent and old search behaviors \cite{vu2015temporal, White2010Predicting}. No significant difference was observed when using such temporal weightings \cite{bennett2012modeling}. However, the influence of a previous query depends on more complex factors than merely the time. A highly related previous query may continue to have a strong influence on the interpretation of the current query in the same session even after a long time span. In this paper, we will use a deep learning framework to account for such factors in the sequential influence patterns.

When building the user profile, another natural intuition is that previous queries and sessions are not always useful \cite{Vu2014Improving}. For example, the profile constructed for previous query "peanut allergy symptoms" is probably useless when the user is searching for "JAVA book". In contrast, if the user once searched "JAVA runtime environment", then we can infer that the book is about "JAVA programming" instead of "JAVA Island". Based on such observations, we also devise a query-aware attention mechanism to learn different weights for previous sessions corresponding to the input query. By doing so, we are able to build a dynamic user profile which attends to the previous sessions that are more important to the current query.

User preference evolves over time, and some previous studies attempted to capture this variation by distinguishing short- and long-term user preferences \cite{bennett2012modeling, li2007dynamic, White2010Predicting}. A short-term user profile is built from recent interactions within the same search session and is useful to predict the following intents in the current search task \cite{white2009predicting, White2010Predicting}. A long-term user profile describes more long-standing user characteristics and is less sparse than a short-term profile \cite{matthijs2011personalizing}, and it is built from previous sessions. Motivated by the hierarchical structures contained in the click-through data, we propose to use hierarchical recurrent neural network as our fundamental framework to model a longer dependency in the sequential data.

More specifically, a short-term user profile is constructed from the user behaviors in the same session to reflect the current search interest. This is implemented by a RNN at low-level. In addition, the long-term user profiles are built from the past sessions, which are designed to reflect the user's long-term interests. This is created using a higher-level RNN in our model. In order to use the long-term user profiles, we attribute weights to each long-term interest profile using an attention model based on the current query. Then the linear combination of all hidden state vectors is produced as a more precise long-term interest vector. Finally, we compute the similarity score between interest vectors and documents as our personalized scores to tailor the original ranking. The whole framework is trained through a learning-to-rank approach, LambdaRank algorithm \cite{burges2005learning}. Taking advantage of the two interest vectors, we can further enhance search results personalization.

In sum, our main contributions are twofold: First, we account for the sequential information contained in click-through data and generate a better user profile through a hierarchical recurrent neural network; second, we apply attention mechanism to scrutinize all previous sessions and highlight the more important sessions dynamically according to present information need.

The rest of paper is organized as follows. Section \ref{sec:related} summarizes previous studies that are related to our paper. The proposed framework to mine sequential information from query log is described in Section \ref{sec:model}.  We discuss experimental setups in Section \ref{sec:experiment}, and analyze the results in Section \ref{sec:results}. We conclude the work in Section~\ref{conclusion}.

\section{Related Work}
\label{sec:related}
The related work to this paper principally concerns three fields: (1) Search Results Personalization, (2) Deep Learning in Information Retrieval and (3) The Applications of Hierarchical Recurrent Neural Network.

\paragraph{Search Results Personalization}
Users' past search interactions with search engine have been revealed beneficial to web search \cite{cai2014personalized}. However, the improvement of ranking quality personalization brings to users mostly depends on the richness of user profiles. Among all intriguing research on personalizing search results, great majority focused on building user profile based on click-through data, since it is both accessible and informative. The personalized features are extracted from click-through data and can be simply categorized into two groups: click features and topical features. 

Users often use search engine to navigate for the same document which was satisfactory under the previously issued queries, and this fact can be used as a safe and low-risk approach in personalization. Teevan et al. \cite{teevan2011understanding} have demonstrated that there is a rich opportunity to personalize search results through recognizing personal navigations. Many subsequent studies \cite{bennett2012modeling,li2014deep,volkovs2015context,white2013enhancing} regarded click-based features as their basic features and investigated more from other angles. It is widely known that personalization sometimes plays against the goal of improving results. Therefore, click entropy is often considered to ensure the expected utility of personalization on certain queries \cite{dou2007large, teevan2008personalize}. More studies were focused on building an appropriate latent topical user profiles. Early studies \cite{bennett2010classification, sieg2007web, white2013enhancing} attempted to build user profiles with topics of clicked documents which are learned from a manual on-line ontology, such as the Open Directory Project (ODP)\footnote{https://dmoztools.net/}. The potential problem of this approach is that some classes of documents may not appear in the on-line ontology, which could limit its application to dealing with new documents \cite{Carman2010Towards, harvey2013building}. Recent studies applied a latent topic model to determining these topics \cite{Carman2010Towards, harvey2013building, vu2015temporal, vu2017search}. 
Besides, topic entropy is also proposed to measure the topical ambiguity of different queries \cite{sontag2012probabilistic}. Li et al. \cite{li2014deep} utilized semantic features produced by a deep learning model as semantic features to improve in-session contextual ranking. In addition to extracting features from query logs, other information can be utilized to improve search results as well. Bennett et al.\cite{bennett2011inferring} incorporated user location-based features into personalization model. Collins-Thompson et al.\cite{collins2011personalizing} studied and evaluated personalization by different reading levels. Different from the previous studies which simply aggregate the historical user behaviors as user profiles, our intuition is that the past behaviors should be differently conducive to current search according to, for example, the time span and the current information need.

Strategies to implement personalization also varied considerably from model to model. Wang et al. \cite{Wang2013Personalized} and Song et al. \cite{song2014adapting} investigated adapting a generic ranking model for personalized search through updating parameters for individual users. The model proposed by Matthijs et al. \cite{matthijs2011personalizing} built user profiles with terms extracted from browsing history, and personalized the rankings with the matching between search snippets and these terms. As for the models focusing on topical user profile with Latent Dirichlet Allocation (LDA) \cite{Carman2010Towards, harvey2013building, vu2015temporal, vu2017search}, the authors directly calculated the similarity between each document and established user profile, and used the similarities as personalized scores to re-rank the results. 
Another common approach is to train a ranking model using the LambdaMART \cite{wu2008ranking} learning algorithm as in \cite{bennett2012modeling, volkovs2015context, white2013enhancing}. LambdaMART is evolved from LambdaRank \cite{burges2005learning} and can distinguish the importance of features automatically based on boosted regression tree. Different from aforementioned approaches, we employ a deep learning framework to personalize search results. Instead of highly depending on the manual features, the framework does not need specific preparation and only takes raw queries and documents with one hot representation of words as inputs.

\paragraph{Deep Learning in Information Retrieval}
Deep learning has been introduced in many information retrieval tasks because of its several advantages, such as the ability to learn word embeddings automatically and to train end-to-end \cite{onal2018neural,mitra2017neural}. 
Ad-hoc rankings have benefited enormously from deep learning methods and great results have been achieved. DSSM \cite{huang2013learning} and CDSSM \cite{shen2014latent} embedded the query-document pair into a semantic space and ranked the results by the similarity between the embeddings of documents and queries. Severyn and Moschitti \cite{severyn2015learning} also used a convolutional deep neural network to represent query-document pair and computed their semantic similarity. The model developed by Palangi et al. \cite{palangi2016deep} shared a similar idea but addressed the embeddings of queries and documents using Long Short-Term Memory (LSTM) \cite{hochreiter1997long}. More recently, some elaborate models like DRMM \cite{guo2016deep} and K-NRM \cite{XiongDCLP17} were proposed to model word-level similarities and achieved even better results. But little has been studied on applying deep learning to search results personalization. Song et al. \cite{song2014adapting} adapted a global ranking model with continue-train for each individual user to personalize the results. Li et al. \cite{li2014deep} improved in-session contextual results by involving semantic features generated by deep learning models. Different from these studies which incorporate deep learning only as a component, we intend to train a complete deep learning framework to personalize the search results. 

\paragraph{The Applications of Hierarchical Recurrent Neural Network}
The hierarchical recurrent neural network (HRNN) was firstly proposed by El Hidi and Bengio \cite{el1996hierarchical} aiming at modeling longer dependence in the sequential data. Since then people gradually realized that many real tasks contain hierarchically sequential dependencies, and HRNN attracts increasing attention from various communities. Quadrana et al. \cite{QuadranaKHC17} employed HRNN to deal with cross- and in-session commodity information in a session-based recommendation. Besides, hierarchical recurrent encoder-decoder was designed to handle text generation, such as building dialogue system \cite{serban2016building} and query suggestions \cite{sordoni2015hierarchical}. In addition to basic hierarchical recurrent neural network, in this paper we also employ an attention model to generate dynamic user profile based on current query.

\begin{figure*}[!ht]
  \includegraphics[width=0.95\linewidth]{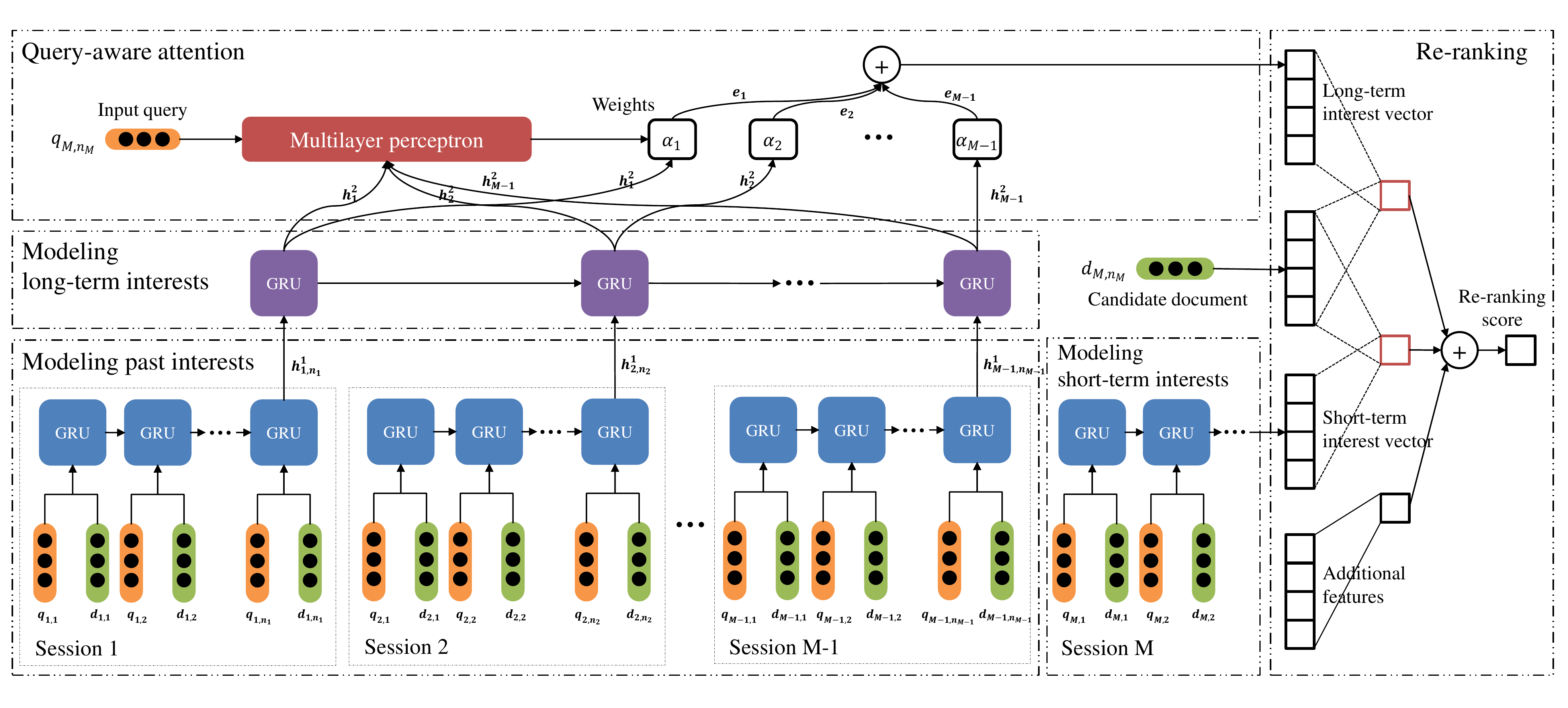}
  \caption{The architecture of our personalization framework. Given query and document representations in each past session, the interest vector of the session is calculated as the last latent state vector of a low-level RNN, which is then proceeded into a high-level RNN. Next, weights are applied on each latent state vector of the high-level RNN according to the input query and generate the long-term interest vector. By matching the interest vectors with candidate documents, and involving additional features, we compute the final personalized scores. }
  \label{fig:model}
\end{figure*}

\section{Personalization Framework}
\label{sec:model}
The one-size-fits-all retrieval models are known to be suboptimal and can be potentially improved with search results personalization. Through re-ranking the unpersonalized results for different users according to their interests, search engines can provide better search experiences to individual users. In this paper, we focus on mining sequential information and building dynamic user profiles using users' past interactions. Specifically, we employ a framework based on hierarchical recurrent neural network with query-aware attention to learn short- and long-term user profiling automatically. The short-term interest is collected from current session, and the long-term interest describes the preferences inferred by other sessions before the current session.

To start with, we formulate our problem as follows. Suppose that for each user \begin{math}u\end{math}, there is a query log \begin{math}\mathcal{L}_u\end{math} which includes past sessions \begin{math}\mathcal{L}_u=\{\mathcal{S}_1,...,\mathcal{S}_{M-2},\mathcal{S}_{M-1}\}\end{math} and current session $\mathcal{S}_{M}$, where \begin{math}M\end{math} is the index of the current session. Each session is defined as a sequence of queries and a list of documents retrieved for each query, denoted as \begin{math} \mathcal{S}_m=\{\{\bm{q}_{m,1},\bm{d}_{m,1,1},...\},...,\{\bm{q}_{m,n_m},\bm{d}_{m,n_m,1},...\}\} \end{math}, where \begin{math}n_m\end{math} represents the total number of queries in the session \begin{math}\mathcal{S}_m\end{math}. In the current session \begin{math}\mathcal{S}_M\end{math}, the user is issuing a query \begin{math}\bm{q}_{M,n_M}\end{math} and a document ranking list \begin{math}\mathcal{D}=\{\bm{d}_1,\bm{d}_2,...\}\end{math} is returned by the search engine, where \begin{math}\bm{d}_i\end{math} is short for \begin{math}\bm{d}_{M,n_M,i}\end{math}. For each document, we want to compute a relevance score separately based on the issued query \begin{math}\bm{q}_{M,n_M}\end{math}, and past interactions \begin{math}\mathcal{L}_u\end{math} and \begin{math}\mathcal{S}_M\end{math}:
\begin{equation*}
\begin{split}
  \text{score}(\bm{d}_i)&=\text{score}(\bm{d}_i|\bm{q}_{M,n_M},\mathcal{S}_M,\mathcal{S}_{M-1},...,\mathcal{S}_1)\\
  &=\text{score}(\bm{d}_i|\bm{q}_{M,n_M})+\text{score}(\bm{d}_i|\mathcal{L}_u)+\text{score}(\bm{d}_i|\mathcal{S}_M),
\end{split}
\end{equation*}
where \begin{math}\text{score}(\bm{d}_i|\bm{q}_{M,n_M})\end{math} represents the relevance between document and query, \begin{math}\text{score}(\bm{d}_i|\mathcal{L}_u)\end{math} and \begin{math}\text{score}(\bm{d}_i|\mathcal{S}_M)\end{math} correspond to the relevance with regard to long- and short-term user interests respectively. Finally, we re-rank the document list \begin{math}\mathcal{D}\end{math} according to the overall score \begin{math}\text{score}(\bm{d}_i)\end{math}, to produce a personalized ranking list. 

As shown in Figure \ref{fig:model}, we devise a deep learning framework based on HRNN to calculate the above personalized scores using query logs and to predict the optimal document ranking list. We elaborate our model based on three main components as follows: (1) Exploiting sequence-enhanced user interests, (2) Modeling dynamic user profiles and (3) Re-ranking.

\subsection{Exploiting Sequence-enhanced User Interests} 
As we stated in Section \ref{sec:intro}, traditional methods for building topical user profiles have several potential problems. Most of them ignored the sequential information contained in historical user behaviors. Inspired by the ability of RNN to model sequential data and the hierarchical structure of query logs, we propose to use a Hierarchical RNN to model user interests with click-through data.

Recurrent neural network has been widely used in natural language processing because of its ability to model the sequences of words and sentences. After inputs at each time point, the RNN cell is able to determine what information will be kept and passed to next state, and what to discard. More sophisticated RNN cells such as Gated Recurrent Unit (GRU) \cite{ChoMGBBSB14} and Long Short-Term Memory (LSTM) \cite{hochreiter1997long} were proposed to solve vanishing gradient problems and learn a long-term dependency. Specifically, we employ GRU as a basic RNN cell in our model, since it has a relatively simpler structure which is easier to train. The reasons we devise a two-level structure include: (1) The click-through data is intrinsically hierarchical with intra-session interactions conveying short-term interest and inter-session interactions reflecting long-term preference; (2) To enhance the ability of our framework to model longer dependency, specifically to learn longer user interests.

\subsubsection{Modeling short-term user interests}
Within a session \begin{math}\mathcal{S}_m\end{math}, a user issues a series of queries. Sometimes this user issues a query at first and reformulates it in the following queries. During this procedure, the user probably clicks on certain documents and may be satisfied by their content. Intuitively, both the satisfactory documents and the order of these queries are highly informative to inferring the real intent and should be utilized while tailoring the ranking of the result lists. 

In our model, the inputs to the low-level RNN is extracted from a subset of the session \begin{math}\mathcal{S}^{'}_{m}=\{\{\bm{q}_{m,1},\mathcal{D}_{m,1}\},...,\{\bm{q}_{m,n_m},\mathcal{D}_{m,n_m}\}\}\end{math}, where \begin{math}\mathcal{D}_{m,i}\end{math} represents the set of SAT-clicked documents under query \begin{math}\bm{q}_{m,i}\end{math}. And we use the concatenation of query vector \begin{math}\bm{q}_{m,i}\end{math} and the average vector \begin{math}\bm{d}_{m,i}\end{math} of document vectors in \begin{math}\mathcal{D}_{m,i}\end{math} to feed the RNN. Note that the average document vector \begin{math}\bm{d}_{m,i}\end{math} is assigned to a zero vector if the user is not satisfied by any documents. To be specific, for each query \begin{math}\bm{q}\in\mathbb{R}^{d_e}\end{math} and document \begin{math}\bm{d}\in\mathbb{R}^{d_e}\end{math}, their representations are calculated as the weighted average of the word representations \begin{math}\bm{w}\in\mathbb{R}^{d_e}\end{math} multiplied by TF-IDF weights, while the words are represented by distributional vectors looked up in an embeddings matrix \begin{math}\bm{W}\in\mathbb{R}^{|V|\times{d_e}}\end{math}. This part could be replaced by a more complex model such as Doc2Vec, CNN, or RNN model, which also involves more parameters and increase the difficulty of training. Then the session-level RNN calculates a series of latent state vectors according to the inputs and previous state vectors, formally:
\begin{equation*}
  \bm{h}_{m,n}^{1}=f(\bm{h}_{m,n-1}^{1},\bm{q}_{m,n},\bm{d}_{m,n})
\end{equation*}
where \begin{math}\bm{h}_{m,n}^{1}\in\mathbb{R}^{d_{h^{1}}}\end{math} and \begin{math}\bm{h}_{m,0}^{1}\end{math} is initialized by a zero vector. Note that we use superscript 1 and 2 to distinguish the notations in low- and high-level RNN respectively such as hidden state vectors. The RNN cell \begin{math}f(\cdot)\end{math} can be vanilla, LSTM, and GRU cell, which is implemented as GRU in our model:
\begin{equation*}
\begin{split}
  \bm{r}_{m,n}^{1}&=\sigma(\bm{W}_{r}^{1} [\bm{q}_{m,n};\bm{d}_{m,n}]+\bm{V}_{r}^{1} \bm{h}_{m,n-1}^{1}),\\
  \bm{z}_{m,n}^{1}&=\sigma(\bm{W}_{z}^{1} [\bm{q}_{m,n};\bm{d}_{m,n}]+\bm{V}_{z}^{1} \bm{h}_{m,n-1}^{1}),\\
  \bm{c}_{m,n}^{1}&=\text{tanh}(\bm{W}^{1} [\bm{q}_{m,n};\bm{d}_{m,n}]+\bm{V}^{1}(\bm{r}_{m,n}^{1}\cdot \bm{h}_{m,n-1}^{1})),\\
  \bm{h}_{m,n}^{1}&=(1-\bm{z}_{m,n}^{1})\cdot \bm{h}_{m,n-1}^{1}+\bm{z}_{m,n}^{1}\cdot \bm{c}_{m,n}^{1},
\end{split}
\end{equation*}
where \begin{math}\bm{h}_{m,n}^{1}\in\mathbb{R}^{d_{h^{1}}}\end{math}, the reset gate \begin{math}\bm{r}_{m,n}^{1}\end{math} and update gate \begin{math}\bm{z}_{m,n}^{1}\end{math} control the trade-off between previous states and present inputs, \begin{math}\sigma(\cdot)\end{math} is the sigmoid function, parameters \begin{math}\bm{W}_{r}^{1}, \bm{V}_{r}^{1}, \bm{W}_{z}^{1}, \bm{V}_{z}^{1}, \bm{W}^{1}\end{math} and \begin{math}\bm{V}^{1}\end{math} are shared across all sessions and updated during training. Each latent hidden state \begin{math}\bm{h}_{m,n}^{1}\end{math} models the short-term interest in session \begin{math}S_m\end{math} after issuing query \begin{math}\bm{q}_{m,n}\end{math}. The dense vector of the last latent hidden state, denoted by \begin{math}\bm{h}_{m,n_m}^{1}\end{math}, is viewed as a representation of the whole session, i.e. the short-term interest vector, which is used as \begin{math}\bm{h}_{m}^{1}\end{math} for short in the following parts.

\subsubsection{Modeling long-term user interests}
Looking at the whole search history, some interests do not change frequently alongside time. For example, if a user kept searching "JAVA runtime environment", "JAVA data type", and "JAVA string to int", the user was probably a programmer. Even though after a very long time this user issued another query "JAVA Book", there should still be a great chance that the user is looking for a book of JAVA programming instead of JAVA Island. Therefore, the parameters of such cell states of interests should change less frequently than the parameters of short-term model, and this fact motivates us to build a hierarchical structure in our framework. Therefore, we add another RNN to encode a higher level information flow.

The high-level RNN takes the short-term interest vectors (for past sessions) \begin{math}\{\bm{h}_1^{1},...,\bm{h}_{M-1}^{1}\}\end{math} outputted by the low-level RNN as inputs. It computes the sequence of user representations \begin{math}\{\bm{h}^{2}_1,..., \bm{h}^{2}_{M-1}\}\end{math} at the end of every session. Formally:
\begin{equation*}
  \bm{h}^{2}_m=f(\bm{h}^{2}_{m-1},\bm{h}_{m}^{1}),
\end{equation*}
where \begin{math}\bm{h}^{2}_m\in\mathbb{R}^{d_{{h}^{2}}}\end{math}, and \begin{math}\bm{h}^{2}_0\end{math} is initialized by a zero vector. The RNN cell \begin{math}f(\cdot)\end{math} is similarly defined in the low-level RNN:
\begin{equation*}
\begin{split}
  \bm{r}^{2}_{m}&=\sigma(\bm{W}^{2}_{r}\bm{h}^{1}_{m}+\bm{V}^{2}_{r} \bm{h}^{2}_{m-1}),\\
  \bm{z}^{2}_{m}&=\sigma(\bm{W}^{2}_{z}\bm{h}^{1}_{m}+\bm{V}^{2}_{z} \bm{h}^{2}_{m-1}),\\
  \bm{c}^{2}_{m}&=\text{tanh}(\bm{W}^2\bm{h}^{1}_{m}+\bm{V}^{2}(\bm{r}^{2}_{m} \cdot \bm{h}^{2}_{m-1})),\\
  \bm{h}^{2}_m&=(1-\bm{z}^{2}_{m})\cdot \bm{h}^{2}_{m-1}+\bm{z}^{2}_{m}\cdot \bm{c}^{2}_{m},
\end{split}
\end{equation*}
where parameters \begin{math}\bm{W}^{2}_{r}, \bm{V}^{2}_{r}, \bm{W}^{2}_{z}, \bm{V}^{2}_{z}, \bm{W}^{2},\end{math} and \begin{math}\bm{V}^{2}\end{math} are shared across all users and updated during training. Analogously, each state \begin{math}\bm{h}^{2}_m\end{math} of this layer models long-term interest of user \begin{math}u\end{math} after the session \begin{math}\mathcal{S}_m\end{math} ends. A simple way to obtain the long-term user interest vector is to adopt the last latent state vector \begin{math}\bm{h}^{2}_{M-1}\end{math} directly. However, this could introduce unnecessary noise as we stated in Section \ref{sec:intro}. To avoid this issue, we implement an attention model to assign query-aware weights to long-term interest vectors \begin{math}\{\bm{h}^{2}_1,...,\bm{h}^{2}_{M-1}\}\end{math} in the end of different sessions, so as to highlight different parts of historical search behaviors dynamically. This intuitive idea is proved effective and better than plain RNN model in our experiments.

\subsection{Building Dynamic User Profiles}
As discussed in Section \ref{sec:intro}, the same past interactions could contribute differently to personalized rankings under different search circumstances. We believe that discriminating previous behaviors is necessary in building an effective user profile. Vu et al. built user profiles using dynamic group formation based on input query \cite{Vu2014Improving}. In our model, we deploy an attention model to apply different weights for each previous session based on current query. Note that in an extreme case, the weights for some interactions should be very high when a user issues a historical query to re-find the information they have searched before. Such re-finding problem has been well studied by Teevan et al. as personal navigation \cite{teevan2011understanding} and is a special case naturally incorporated in our model.

Attention mechanism was first proposed in machine translation to deal with the limitation of fixed-length representation of input sentences \cite{bahdanau2014neural}. Similarly we hypothesize that encoding the click through data into a fixed-length representation is not reasonable when we are personalizing search results under different situations. Therefore, with respect to the current query, we assign query-aware weights to past interactions and compute a more precise long-term user profile. In other words, we suppose that the long-term interest vector is influenced by the current information need dynamically. At the end of session \begin{math}\mathcal{S}_{M-1}\end{math}, we calculate weights \begin{math}\{\alpha_1,...,\alpha_{M-1}\}\end{math} for each past interest vector in \begin{math}\{\bm{h}^{2}_1,...,\bm{h}^{2}_{M-1}\}\end{math} as follow.

\begin{equation*}
\begin{split}
  e_i&=\phi(\bm{q}_{M,n_M},\bm{h}^{2}_i),\\
  \alpha_i&=\frac{\text{exp}(e_i)}{\sum_{j=1}^{M-1}\text{exp}(e_j)},
\end{split}
\end{equation*}

where \begin{math}\phi(\cdot)\end{math} is a multilayer perceptron (MLP) with \begin{math}\tanh(\cdot)\end{math} as activation function in our model, which is updated during training and could be replaced by more complex functions in the future. Then the query-aware long-term interest vector \begin{math}\bm{h}_{M-1}^{2,q}\end{math} is computed by a weighted linear combination of \begin{math}\{\bm{h}^{2}_1,...,\bm{h}^{2}_{M-1}\}\end{math}: 

\begin{equation*}
  \bm{h}^{2,q}_{M-1}=\sum_{i=1}^{M-1}\alpha_i\bm{h}^{2}_i.
\end{equation*}

In sum, we denote the final short- and long-term interest vectors as $\bm{h}^{1}_{M,n_m}$ and $\bm{h}^{2,q}_{M-1}$ respectively and use them to re-rank the search results in the following.

\subsection{Re-ranking}
Finally we re-rank the original results using the personalized information we collect. Given the short- and long-term interest vectors, $\bm{h}^{1}_{M,n_m}$ and $\bm{h}^{2,q}_{M-1}$, we calculate the personalized ranking scores of a document by measuring its similarity to these two interest vectors:
\begin{equation*}
\begin{split}
  \text{score}(\bm{d}_i|\mathcal{L}_u)&=\text{sim}((\bm{h}^{2,q}_{M-1})^T \bm{W}_L, \bm{d}_i),\\
  \text{score}(\bm{d}_i|\mathcal{S}_M)&=\text{sim}((\bm{h}^{1}_{M,n_m})^T \bm{W}_S, \bm{d}_i),
\end{split}
\end{equation*}
where \begin{math}\bm{W}_S\in\mathbb{R}^{d_{h^1}\times d_e}, \bm{W}_L\in\mathbb{R}^{d_{h^2}\times d_e}\end{math} are two similarity matrices whose functions are to project the interest vectors into the same semantic space as the documents, which are optimized during training. The similarity function \begin{math}\text{sim}(\cdot)\end{math} is defined as:
\begin{equation*}
  \text{sim}(\bm{X}, \bm{Y})=\frac{\bm{X}^T\bm{Y}}{||\bm{X}||\cdot||\bm{Y}||}.
\end{equation*}

In addition to personalization scores calculated by our model, we also incorporate query-document relevance feature and click-based features as additional features. Since the original query-document features are inaccessible in our dataset, here we use the original position of the document as a feature. Also, the click features include the total number of historical clicks on the candidate document by the user, the number of clicks on candidate document under the input query  by the user and the click entropy of input query. The reason we incorporate click entropy is because the value of personalization varies a lot across different queries, and indiscriminately applying personalization on all queries could produce an adverse effect on the overall quality \cite{dou2007large,teevan2008personalize}. Note that as for fair comparison, the baseline model is a more complex model that incorporates more than these three features. And our idea is focused on building topic-based user profile instead of investigating on click-based features as discussed in Section \ref{sec:intro}. These additional features are fed into a multilayer perceptron (MLP) with \begin{math}\tanh(\cdot)\end{math} as the activation function. Finally, we sum up the scores calculated by different parts as the final relevance score. 

We choose a basic ranking algorithm, LambdaRank \cite{burges2005learning}, to train the whole framework. We generate training pairs from query logs by treating the SAT-clicked documents as relevant samples and the others as irrelevant ones. Then we use the representations of document pairs to calculate the loss. Take a pair of relevant document \begin{math}\bm{d}_i\end{math} and irrelevant document \begin{math}\bm{d}_j\end{math} as an example. The loss function is the product of cross entropy between desired probabilities and predicted probabilities and the change of metrics, $\Delta$, while swapping the positions of the two documents, defined as:
\begin{equation*}
  \text{loss}=(-\overline{p}_{ij}\text{log}(p_{ij})-\overline{p}_{ji}\text{log}(p_{ji}))|\Delta|,
\end{equation*}
where \begin{math}p_{ij}\end{math} represents the predicted probability that \begin{math}\bm{d}_i\end{math} is more relevant than \begin{math}\bm{d}_j\end{math}, and \begin{math}\overline{p}_{ij}\end{math} represents the real probability. Specifically, the predicted probabilities are computed by a logistic function,
\begin{equation*}
  p_{ij}=\frac{1}{1+\text{exp}(-(\text{score}(\bm{d}_i)-\text{score}(\bm{d}_j)))}.
\end{equation*}

\section{Experiment Setup}
\label{sec:experiment}

\subsection{Dataset and Evaluation}
The dataset in our experiment is sampled randomly by users from the logs of a commercial search engine, comprising click-through data of the users between \begin{math}1^{st}\end{math} January 2013 and \begin{math}28^{th}\end{math} February 2013. Each piece of data in the log contains an anonymous user identifier, a query, a session identifier, query issued time, the top 20 URLs retrieved by the search engine, clicks and dwelling time. The logs are collected when personalization support was not applied, so that our results are guaranteed not to be biased toward other personalization signals. The basic statistics is shown in Table \ref{tab:dataset}.

\begin{table}[!t]
  \vspace{+0.5cm}
  \caption{Basic statistics of the dataset.}
  \label{tab:dataset}
  \begin{tabular}{cc||cc}
  	\toprule
    Item & Statistic & Item & Statistic \\
  	\midrule
    \#days & 58 & \#distinct queries & 1,624,496\\
  	\#users & 33,204 & \#sessions & 654,776 \\
	\#queries & 2,665,625 & \#SAT-clicks & 1,228,028 \\    
  	\bottomrule
  \end{tabular}
\end{table}

Following \cite{vu2015temporal,bennett2012modeling}, we similarly regard the click that has a dwelling time of more than 30 seconds or is the last one in the session as a satisfied click (SAT-click). We use the first six weeks of data to generate basic user profiles and use the remaining two weeks to train and test the models. Note that we use not only sessions from the first six weeks but all sessions before current session to calculate long-term interest vector. We split the last two-week data into training and test sets according to the sessions instead of the dates. The motivation behind this is that two weeks are a relatively short time span and the distribution of queries sometimes is uneven across dates. Besides, sessions can be viewed as search activities with independent intents of users, thus can be reasonably divided into different sets. For each user, we divide the sessions into 5:1 as training and test set respectively in the time order of sessions and treat the last one fifth sessions of training set as validation set. For each URL in the logs we retrieve its content and we remove the URLs that cannot be found anymore. Then we preprocess the documents by removing stopwords and punctuations. To ensure an effective segmentation of training and test dataset, we also remove the users who had less than 4 sessions. 

We measure the quality of each ranking results using mean average precision (MAP), mean reciprocal rank (MRR), precision@1 (P@1) and average click position (Avg. Click). In addition to these mainstream metrics, we further evaluate the rankings by measuring the actual improvements on inverse document pairs \cite{Joachims2005Accurately}, i.e. a clicked document and the skipped documents ranked before it. Our reason behind this is that clicks are not only based on the relevance of documents but also the positions \cite{Craswell2008AnEC}. This kind of position bias may make mainstream metrics somewhat problematic. For example, the low-ranked documents have lower probabilities to be examined. Even though some of them may be relevant, they are not clicked because they are not examined by users. These results are labeled as irrelevant by traditional metrics. Consequently, if we boost the positions of the unexamined documents, the traditional evaluation metrics cannot reflect the real changes. Therefore, we use the strategy, which considers clicked documents are better than skipped documents, proposed by Joachims et al. \cite{Joachims2005Accurately} to collect inverse document pairs. We compute the number (\#Better) and percentage of improved pairs (P-Improve) in each personalized ranking to evaluate the reliable improvements made by personalization methods. 

\subsection{Baselines}
\label{sub:baseline}
In addition to the original ranking generated by the search engine, which is usually of very high quality, we further reproduce several state-of-the-arts personalization models as follows.

{\bfseries P-Click:} Users often issue an identical query to re-find the previously viewed information. This observation was confirmed by Teevan et al \cite{teevan2011understanding} as a safe opportunity to improve the quality of search engine. Dou et al. proposed P-Click \cite{dou2007large} as a basic personalization strategy. As for a user \begin{math}u\end{math}, P-Click calculates the personalized score of a document \begin{math}d\end{math} as the percentage of clicks on this document from past queries that are identical to the input query \begin{math}q\end{math}, formally:
\begin{equation*}
  \text{score}(d|q,u)=\frac{|\text{clicks}(q,d,u)|}{|\text{clicks}(q,\bullet,u)|+\beta},
\end{equation*}
where \begin{math}|\text{clicks}(q,d,u)|\end{math} represents the click counts on \begin{math}d\end{math} from query \begin{math}q\end{math} by user \begin{math}u\end{math} in the past, \begin{math}|\text{clicks}(q,\bullet,u)|\end{math} is the total click counts on query \begin{math}q\end{math} by user \begin{math}u\end{math}, and \begin{math}\beta\end{math} is a smoothing factor set as 0.5. P-Click re-ranks the results based on this personalized score and combines the personalized ranking with original ranking using Borda' ranking fusion method. This model is also used as a baseline in \cite{matthijs2011personalizing}. Note that this method is purely based on the click information.

Besides, we also care about the quality of user profile built by our model and past studies. Many past studies built user profiles based on clicked documents over a topic space. They assume that the topic that users are interested in is reflected by the topics of documents they clicked \cite{sontag2012probabilistic}. These past studies tailored the original ranking either directly using similarity between user profile and documents \cite{Vu2014Improving} or training a supervised model with diverse features \cite{bennett2012modeling}. However, the intrinsic idea of these methods is the same - using the aggregation of topics of past documents as user profiles.

{\bfseries SLTB:} Bennett et al \cite{bennett2012modeling} implemented a personalization method by extracting diverse features from short- and long-term behavior (SLTB). SLTB generates personalized features through the combination of four options: (1) feature types (click-based or topic-based), (2) document coverage (across all queries; under queries identical/generalizations/specializations to the current query.), (3) temporal angle (historic, session and aggregate) and (4) temporal decay (able or disable). The decay is defined by the function \begin{math}{0.95}^{p(q_r)-1}\end{math}. Here \begin{math}p(q_r)\end{math} refers to the number of queries preceding the current query and \begin{math}0.95\end{math} is a chosen decay factor. As for the topical representation of the documents, it uses the top two levels of the Open Directory Project hierarchy as the 207 labels of documents and trains a classifier to predict the categories of documents. Besides, SLTB also implements click entropy \cite{dou2007large}, topic entropy \cite{sontag2012probabilistic}, and other features. All these features are fed into a learning-to-rank model, LambdaMART \cite{wu2008ranking}, to generate a personalized ranking. As for the parameters used in our experiments, especially those in the Learning-to-rank models, we initialize them with parameters same as in \cite{bennett2012modeling} and then tune them using the validation set.

\begin{table*}[!t]
 \caption{Overall performances of models. \bfseries{Bold} indicates the main model proposed by this paper, which is also the best among all compared models. The improvements achieved by HRNN and HRNN+QA on MAP, MRR, P@1 and Avg.Click are significantly larger than improvements made by any baseline model with paired t-test at p-value<0.01.}
  \label{tab:evaluation}
  \begin{tabular}{cccccccc}
  	\toprule
  	Model & MAP & MRR & P@1 & Avg. Click & \#Better & P-Improve\\
    \midrule
  	Original Ranking & .7226 & .7334 & .5931 & 2.292 & - & -\\ 
    P-Click & .7348 & .7467 & .6015 & 2.138 & 4,834 & .1419\\
    PTM & .6679 & .6801 & .5244 & 2.578 & 9,684 & .2845\\
	SLTB & .7776 & .7881 & .6698 & 2.054 & 16,257 & .4777\\ 
	SLTB+PTM & .7830 & .7929 & .6716 & 1.998 & 16,602 & .4878\\ \hline
    HRNN & .7989 & .8107 & .7039 & 1.930 & 18,166 & .5338\\  
    HRNN+QA & \bfseries{.8017} & \bfseries{.8135} & \bfseries{.7067} & \bfseries{1.904} & \bfseries{18,609} & \bfseries{.5468}\\
    \bottomrule
  \end{tabular}
\end{table*}

{\bfseries PTM:} Learning document representations from manual labels is problematic according to Carmen \cite{Carman2010Towards}, since the categories of some documents are missing in the ontology. Instead, they proposed to use an unsupervised approach to learn a multinomial distribution for documents on the latent topics, and this method is also used by Vu et al. \cite{Vu2014Improving} later. Here we reproduce the PTM model proposed by \cite{harvey2013building} as our baseline. PTM calculated personalized scores based on the likelihood of documents given both query and the user as:
\begin{equation*}
  \text{score}(d|q,u) \propto P(d) \prod_{w \in q}\sum_z P(w|z)P(u|z)^{\lambda}P(z|d),
\end{equation*}
where \begin{math}\lambda\end{math} balances the weight of user's topical interest influencing on the overall ranking, and \begin{math}P(d)\end{math} is estimated with Dirichlet smoothing based on the relative frequency of clicks on \begin{math}d\end{math} in the whole log:
\begin{equation*}
  \widehat{P}(d) = \frac{\#\text{clicks}(d)+\sigma \frac{1}{|\mathcal{D}|}}{\sum_{d_i}\#\text{clicks}(d_i)+\sigma}.
\end{equation*}

{\bfseries SLTB+PTM:} We also replace the topical features in SLTB by the features generated by the topic model from PTM, and keep the other features the same as in SLTB. We use this method as the fourth baseline.

\subsection{Our Models}
To verify the effects of our personalization framework, we train two personalization models formed with different compositions but under a same learning setting. Specifically:

{\bfseries Hierarchical RNN+Query-aware Attention (HRNN+QA):} The complete model stated in Section \ref{sec:model};

{\bfseries Hierarchical RNN (HRNN):} The Query-aware Attention is disabled and the long-term interest vector is represented by the last latent vector of the high-level RNN layer;

To determine a group of appropriate parameters, we apply a grid search according to the performance of the model on the validation set. We empirically use a two-layer network for the attention and additional MLP. We use a range of word embedding sizes \begin{math}d_e\in\{300, 1000\}\end{math}, sizes of short-term interest vector \begin{math}d_{s1}\in\{100, 200, 300, 500\}\end{math}, sizes of long-term interest vector \begin{math}d_{s2}\in\{200, 400, 600, 1000\}\end{math}, number of hidden units in attention MLP \begin{math}d_{a}\in\{512, 1024\}\end{math}, number in additional MLP \begin{math}d_{f}\in\{32, 64, 128\}\end{math}, and learning rates \begin{math}\lambda\in\{1e^{-4},1e^{-3},1e^{-2}\}\end{math}. No noticeable differences are observed while changing learning rates. When embedding size is larger, a relatively smaller size of RNN state will yield worse results. Considering the balance of efficiency and result quality, we finally chose a combination of parameters as \begin{math}d_e=300, d_{s1}=300, d_{s2}=600, d_{a}=1024, d_{f}=64\end{math} and \begin{math}\lambda=1e^{-3}\end{math}. 

Similar to \cite{severyn2015learning}, we initialize the word embedding matrix \begin{math}W\end{math} with a pre-trained unsupervised model and keep it fixed during the training. In this experiment, we train an embedding matrix on the documents from training dataset using the Google word2vec tool \footnote{https://code.google.com/archive/p/word2vec/}.
Finally, as we state at Section \ref{sec:model}, we use Mean Average Precision as our metric to calculate $\Delta$ in the learning-to-rank model, since MAP encourages to rank the most relevant documents at top positions, but we still evaluate the final rankings with four different metrics. Besides, we adopt an early stop strategy to end the training when the average loss on validation set stops decreasing in three continuous epochs. 

\section{Experimental Results and Analysis}
\label{sec:results}
As discussed in Section \ref{sec:intro}, the sequences of previous queries and sessions are valuable to personalization and a static user profile is not enough. Therefore, we are very interested in two research questions: (1) Is hierarchical recurrent neural network able to mine sequential information from query logs and learn a better topical user profile? (2) Does attention mechanism help to highlight different parts of search histories dynamically? To answer the questions, we firstly evaluate the overall performances of models stated in the previous section. Then we will discuss the effectiveness of attention mechanism by visualizing the provided weights. In addition, we will analyze the performances of baseline models and our models under different situations, including on queries with different click entropies, on repeated and non-repeated queries, and on queries at different positions in a session. 

\subsection{Overall Performance}
We evaluate the results generated by the search engine, baseline models and our proposed models using MAP, MRR, P@1, Avg. Click, \#Better, and P-Improve. In addition to original ranking, the baseline models include P-Click, PTM, SLTB and SLTB+PTM, and our frameworks include HRNN and HRNN+QA. All the scores are computed over all test queries. Results are shown in Table \ref{tab:evaluation}. We have the following observations:

(1) All personalization methods except PTM successfully improve the original ranking. Directly using user profiles to re-rank the results might cause many problems. For example, applying personalization on queries with low click entropies may increase the risk of personalization. It is observed that P-Click only fixes about half of inverse document pairs of those improved by PTM. In contrast, though very simple, P-Click improves the original ranking in a relatively safe way. SLTB and SLTB-PTM significantly improve original ranking with paired t-test at p<0.01. This proves that a well-trained learning-to-rank model with rich features can effectively yield personalization under different circumstances. 

(2) As shown in Table \ref{tab:evaluation}, our models generate great improvements over all baseline models and the differences are statistically significant with paired t-test at p<0.01 level. More specifically, HRNN+QA obtains rankings that are 0.0187 higher in MAP than the rankings generated by SLTB+PTM. As for reliable improvements, HRNN+QA produces a 12\% increase in the number of improved inverse document pairs than SLTB+PTM. This outcome demonstrates that our models are able to learn more precise user profiles, and consequently yield better personalization. 

(3) We find that attention mechanism works on HRNN and improves its results with 443 more inverse document pairs fixed. However, such improvement is not statistically significant. One possible reason is that the attention model is difficult to apply on a relatively long sequence (around 80 queries were issued per user). Since the attention layer calculates weights that despite the time span, some old behaviors might be assigned undeserved weights in the final profile and destroy the sequential information. One potential replacement of attention model is to apply Reinforcement Learning to discretely select the search sessions and queries, which successfully worked on sentence classification task with noise data \cite{FengHZYZ18}.

The overall performances provide us a general evaluation on these methods. \textbf{Observing the significantly improvement generated by our model, it is safe to draw conclusion that hierarchical recurrent neural network is able to model the sequential information and build better user profiles than traditional methods.} To further analyze the function of attention model introduced in our framework, it is necessary to find out on what queries the model applies a higher weight. In the next section, we give a real example and visualize the weights.

\subsection{Visualization of Weights Assigned by Attention Model}
\begin{figure}[!t]%
\vspace{-0.5cm}
\setlength{\abovecaptionskip}{0.2cm}
\setlength{\belowcaptionskip}{-0.4cm}
\includegraphics[width=0.95\linewidth]{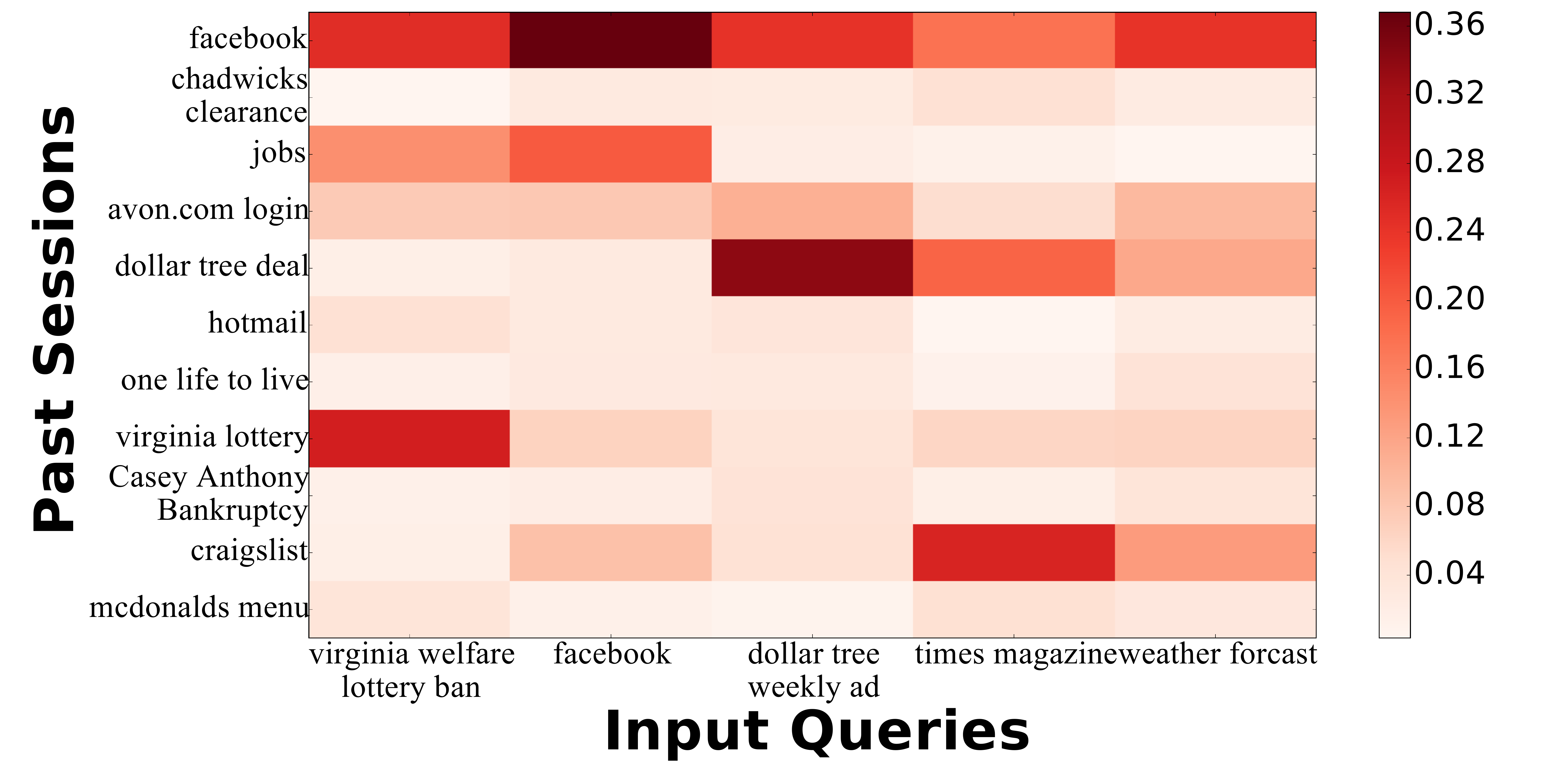}%
\caption{The weights of past sessions when different queries are issued. A darker area indicates a larger weight.}\label{fig:attention}%
\end{figure}%

While personalizing search results in a specific situation, it is unnecessary for search engine to use the whole search history. Depending on the input query, some previous information is useless. Blindly using all data could bring noise into user profiles. Intuitively, sessions that have more similar intents should contribute more in building the current user profile. In this paper, we implement a query-aware attention mechanism to highlight different past interactions on the basis of input query and generate user profile dynamically. Table \ref{tab:evaluation} shows that using this kind of dynamic user profiles improves personalization. To further analyze the influence of attention mechanism, we sample a user from query logs who has relatively rich search history, and visualize the weights applied on different sessions. To make it clearer, we represent the intent of each past session with a typical query from it and sum up the weights of different sessions with the same intent. We select five input queries in the test data and remove the past sessions that have weights less than 0.01 for any of the queries.

As shown in Figure \ref{fig:attention}, we find that the long-term interest vector attends to past sessions that are more relevant to input queries. In general, the attention can filter the irrelevant information in the query logs. For example, when a user issues a query "virginia welfare lottery ban", the past session containing "virginia lottery" gains the largest weight and the other sessions obtain lower weights. We also find that the past sessions for "facebook" are weighted highly in all five input queries, which is contrary to our intuition. One possible reason is that queries on "facebook" have been issued frequently in the past, so the aggregation of these different past sessions leads to a higher weight for "facebook" sessions. Input queries such as "times magazine" and "weather forest" whose topics did not occur in the search history obtain weights that do not vary too much. \textbf{The visualization of weights on past sessions shows that the attention model is able to distinguish the relatively important sessions in the user query logs.}

In order to further verify the effects of our model, it is worthwhile to analyze on what kinds of queries that our methods could achieve larger improvements. In the following sections, we focus on comparing our frameworks with two SLTB baseline models. We use $\Delta$MAP, the change of MAP over original ranking, as the main metric to describe results.

\begin{figure}[!t]
\vspace{-1cm}
\setlength{\abovecaptionskip}{0.2cm}
\setlength{\belowcaptionskip}{-0.4cm}
\includegraphics[width=0.95\linewidth]{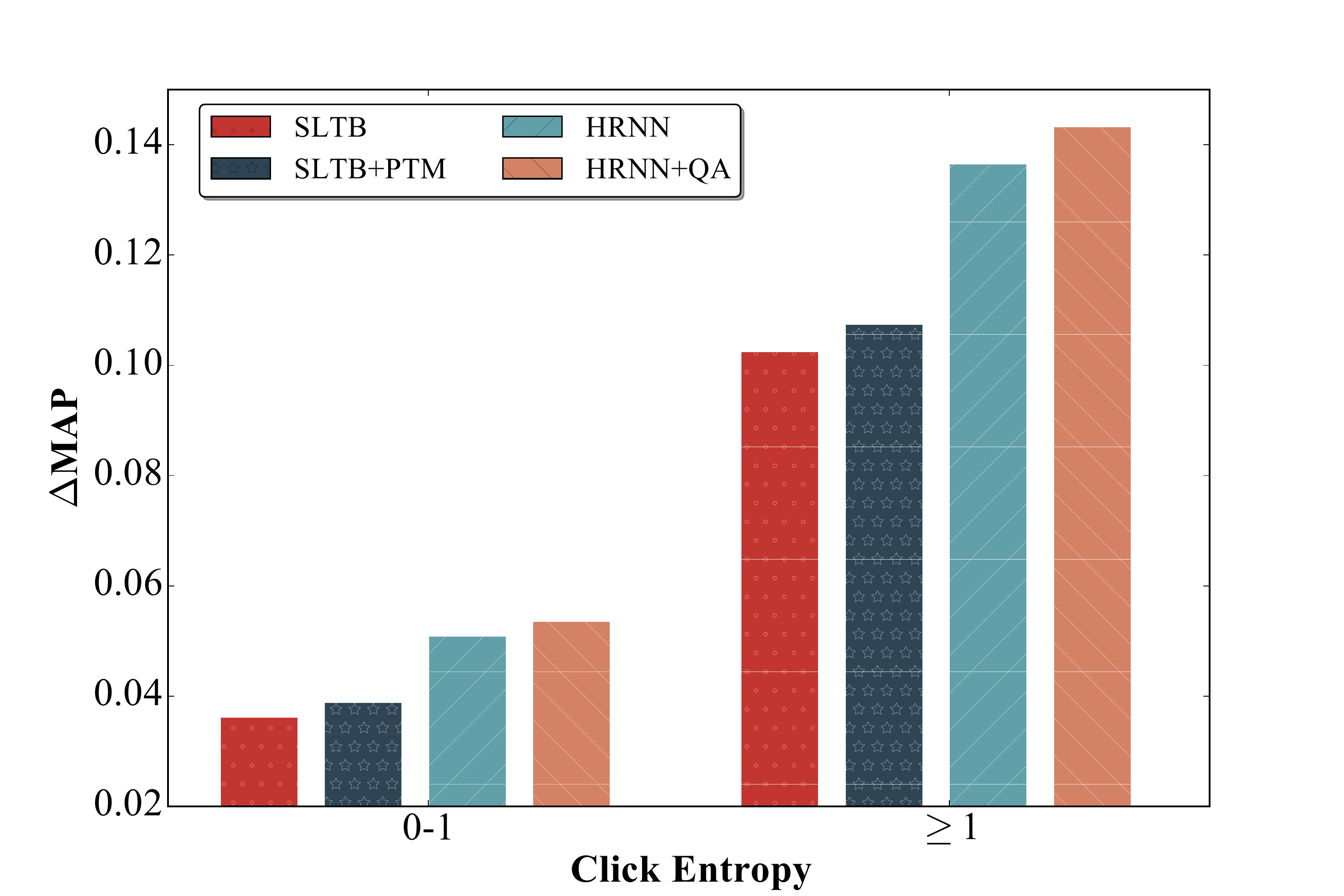}
\caption{The improvements over original ranking on queries with different click entropies.}\label{fig:clickentropy}
\end{figure}

\subsection{Performance on Navigational and Non-navigational Queries}

A larger click entropy often indicates a higher potential for personalization, as it represents a larger probability of diverse user intents \cite{dou2007large,teevan2008personalize}. To study this problem, we firstly group the queries with cutoff of click entropy at 1.0. Teevan et al. \cite{teevan2011understanding} used it as an indicator of distinguishing general navigational queries. Then we compare the improvements made by different models on two groups of queries.

As seen in Figure \ref{fig:clickentropy}, all personalization models produce improved rankings on both groups of queries, and the average improvements on non-navigational queries (click entropy$\geq$1) are much larger than on navigational queries (click entropy<1). In general, our model based on HRNN+QA outperforms any other methods and all of our models outperform the baseline methods (p<0.01). Specifically, as for navigational queries, the $\Delta$MAP of our framework based on HRNN+QA is 0.0124 higher than the best baseline model's (SLTB+PTM). Such improvement further increases to 0.0337 when click entropy is larger than 1.0. Also, all of our two models have improvements of about 0.03 on the non-navigational queries. These results confirm that our framework works better on non-navigational queries than navigational queries. 

\subsection{Performance on Repeated and Non-repeated Queries}

In this experiment, we categorize test queries into two groups: repeated queries and non-repeated queries. A repeated query is a query that the user has issued in the past, and this indicates that the user probably issues the current query to re-find the same information. The click-based personalization methods we discuss in Section \ref{sec:related} highly depend on the information extracted from these repeated queries. If a model never saw the query in the past, most click-based features will be disabled. Therefore, it is worthwhile to evaluate our framework on these non-repeated queries when click based features do not work.

\begin{figure}[!t]
  \vspace{-0.8cm}
  \setlength{\abovecaptionskip}{0.2cm}
  \setlength{\belowcaptionskip}{-0.5cm}
  \includegraphics[width=0.95\linewidth]{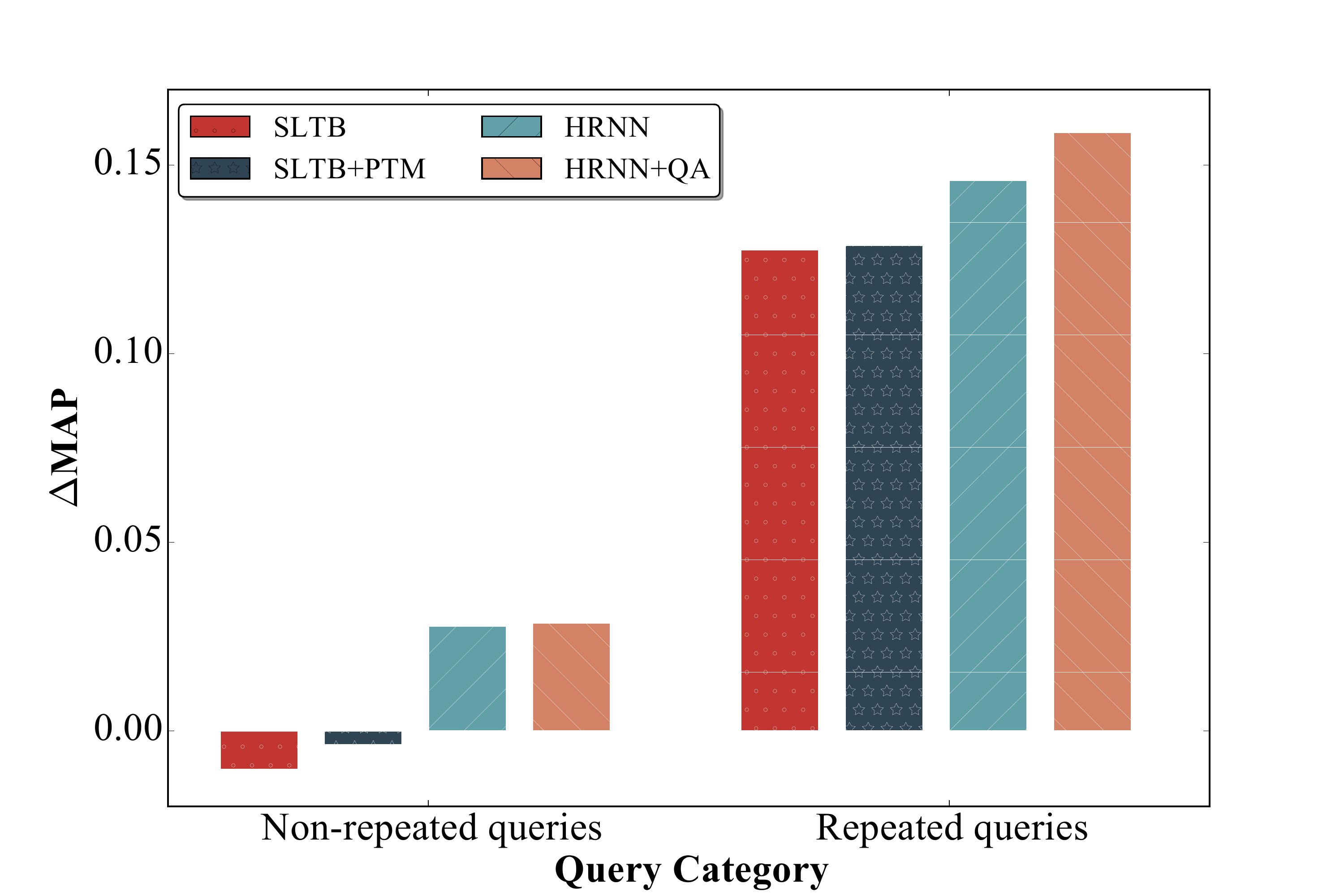}
  \caption{The improvements over original ranking on repeated queries and non-repeated queries.}
  \label{fig:repeat}
\end{figure}

As shown in Figure \ref{fig:repeat}, all methods significantly improve the original results on repeated queries, while our models also outperform the baseline models. The MAP of ranking generated by HRNN+QA on these queries is around 0.03 larger than by SLTB+PTM. With the topical information, the model can better promote the clicked results. Note that personalization on the non-repeated queries is very challenging when baseline models fail improving the search results. On the contrary, our framework successfully generalizes to the queries that are never seen before. Due to the lack of click information for non-repeated queries, only topical features can play a role in tailoring the search results. This observation shows that it is insufficient to simply use an aggregation of clicked documents, and it demonstrates the effectiveness of our model in learning a better user profile. In addition, it shows that applying attention model or not does not make a big difference on non-repeated queries. The majority of the improvements achieved by attention model occurs on repeated queries, which means attention model further highlight the very similar previous queries and reduce the noise.

\subsection{Performances on Queries at Different Positions in Sessions}
In Table \ref{tab:evaluation}, we find that hierarchical structure achieved the best compared with baseline models, but the roles of long- and short-term interest vectors in our framework are still unclear. In this experiment we analyze the performances of each part on queries at different positions in a session. To avoid the influence of additional features, we focus our experiment on non-repeated queries, where click-based features are no more useful. Specifically, we train a plain RNN model whose hierarchical RNN layer is replaced by a one-layer RNN. The model unpacks the session-segmented inputs of HRNN and feeds the whole historical data to the RNN to calculate evolving latent state vectors. In addition, we set a short-term RNN model by disabling the high-level RNN and using the short-term interest vector only, and also a long-term RNN model in the similar way. Note that the long-term RNN model still depends on the low-level RNN to modeling past interests. We analyze the performances of above frameworks and compare them with the complete HRNN. 

\begin{figure}[!t]%
 \vspace{-0.8cm}
 \setlength{\abovecaptionskip}{0.2cm}
 \setlength{\belowcaptionskip}{-0.4cm}
 \includegraphics[width=0.95\linewidth]{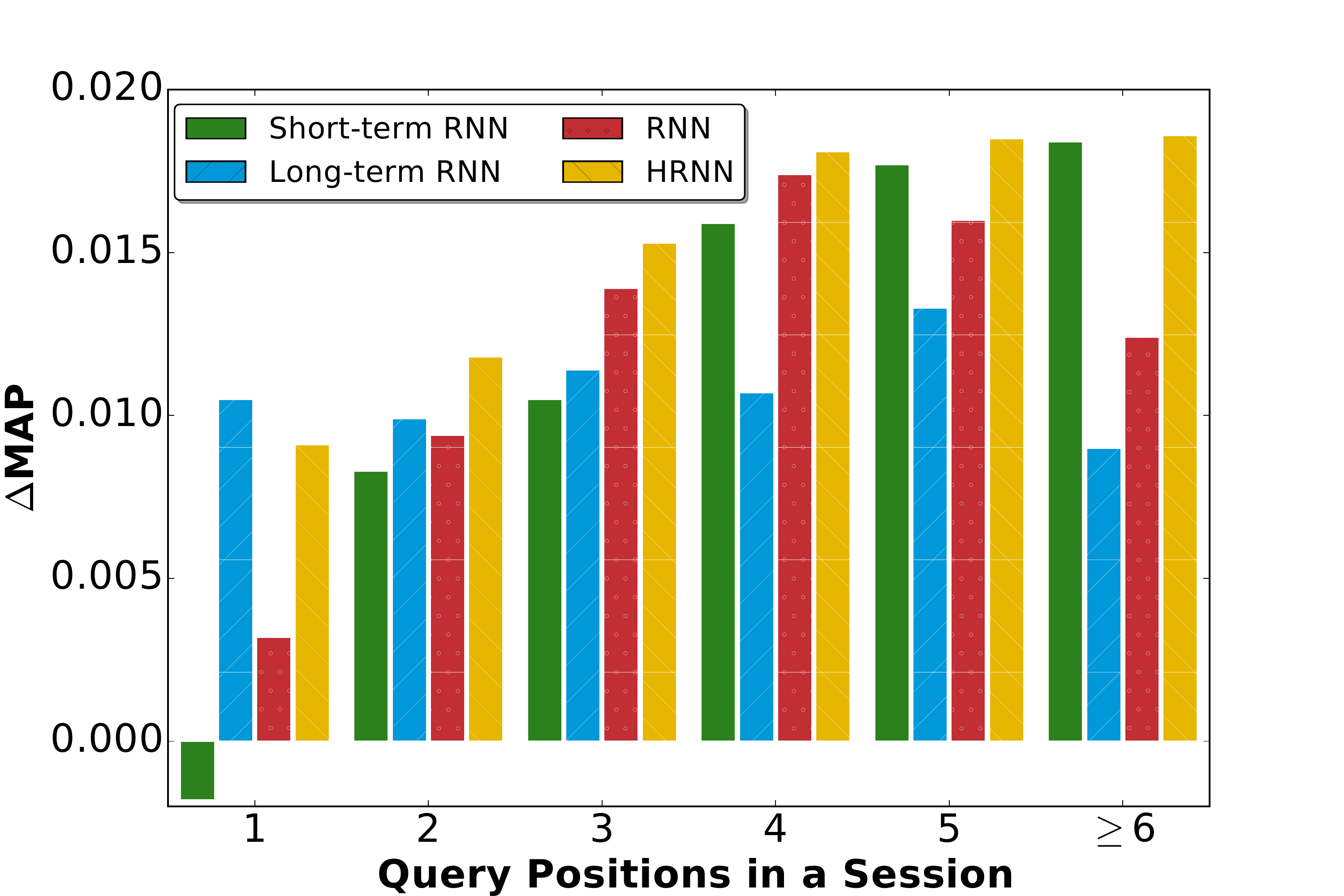}%
 \caption{The improvements of models vs. positions of query in sessions.}\label{fig:querypos}%
\end{figure}

From Figure \ref{fig:querypos}, we find that short-term RNN steadily improves the results more with the increasing of query positions, because more information is available at a higher position during the session. This observation is consistent with Bennett et al. \cite{bennett2012modeling}. In contrast, long-term RNN in our model generates relatively stable benefits on queries at different positions. On the queries at first position in each session, short-term RNN fails to improve the search results, since no session information is available. In that case, the zero short-term vector is used, which turns out to be ineffective. \textbf{By analyzing the queries at first positions, we find that hierarchical structure is better in modeling long-term dependency, and it produces improvements two times larger than plain structure.}

\section{Conclusion}
\label{conclusion}
Large amount of data is generated every single day alongside the interactions between users and search engines. Previous studies have demonstrated that by extracting diverse features from large-scale query logs, search engines are able to tailor the original ranking to satisfy individual users. However, none of these studies successfully exploit the sequential information contained among queries and sessions. In this paper, we propose to apply a deep learning framework to solve this problem. 
Specifically, we deploy a hierarchical recurrent neural network with query-aware attention mechanism to dynamically generate the topical user profiles. The evaluation on the query logs from a commercial search engine shows that our framework significantly outperforms the existing non-deep learning methods. 
In contrast to previous studies which assigned a timespan-based decay on historical data or ignored the sequential information, our model is able to automatically decide what to keep in the user profiles and consequently generate better user profiles. Our experiments also confirm that the query-aware attention model is able to highlight the important parts from previous queries and sessions. 
To further improve our framework, we can replace the attention model with a Reinforcement Learning model, leveraging its ability to discretely select past data. Besides, how to leverage documents skipped by users is also worth further studying.

\begin{acks}
Zhicheng Dou is the corresponding author. This work was supported by National Key R\&D Program of China No. 2018YFC0830703, National Natural Science Foundation of China No. 61872370 and No. 61502501, and the Beijing Natural Science Foundation No. 4162032.
\end{acks}

\bibliographystyle{ACM-Reference-Format}
\bibliography{HSQA} 

\end{document}